\documentclass[journal=jacsat,article]{achemso}

\setkeys{acs}{usetitle = true}
\setkeys{acs}{usetitle = true}

\usepackage[version=3]{mhchem} 
\usepackage{amssymb}
\usepackage{graphicx}
\usepackage{dcolumn}
\setkeys{acs}{usetitle = true}

\author{J. von Vangerow}
\affiliation{Physikalisches Institut, Universit{\"a}t Freiburg, 
79104 Freiburg, Germany}
\author{A. Sieg}
\affiliation{Physikalisches Institut, Universit{\"a}t Freiburg, 
79104 Freiburg, Germany}
\author{F. Stienkemeier}
\affiliation{Physikalisches Institut, Universit{\"a}t Freiburg, 
79104 Freiburg, Germany}
\author{M. Mudrich}
\email{mudrich@physik.uni-freiburg.de}
\affiliation{Physikalisches Institut, Universit{\"a}t Freiburg, 
79104 Freiburg, Germany}
\author{A. Leal}
\affiliation{Departament ECM, Facultat de F{\'i}sica and IN$^2$UB, 
Universitat de Barcelona, 08028 Barcelona, Spain}
\author{D. Mateo}
\affiliation{Departament ECM, Facultat de F{\'i}sica and IN$^2$UB, 
Universitat de Barcelona, 08028 Barcelona, Spain}
\alsoaffiliation{Department of Chemistry and Biochemistry, California State University at Northridge, 18111 Nordhoff St., Northridge, CA 91330, USA}
\author{A. Hernando}
\affiliation{Laboratoire de Chimie Physique Mol{\'e}culaire,
Swiss Federal Institute of Technology Lausanne (EPFL), 1015 Lausanne, Switzerland}
\author{M. Barranco}
\affiliation{Departament ECM, Facultat de F{\'i}sica and IN$^2$UB, 
Universitat de Barcelona, 08028 Barcelona, Spain}
\author{M. Pi}
\affiliation{Departament ECM, Facultat de F{\'i}sica and IN$^2$UB, 
Universitat de Barcelona, 08028 Barcelona, Spain}

\title{Desorption Dynamics of Heavy Alkali Metal Atoms (Rb, Cs) off 
the Surface of Helium Nanodroplets}


\begin{document}
\setkeys{acs}{usetitle = true}

\begin{abstract}
We present a combined ion imaging and density functional theory
study of the dynamics of the desorption process of rubidium  and 
cesium  atoms off the surface of helium nanodroplets upon excitation 
of the perturbed $6s$ and $7s$ states, respectively. 
Both experimental and theoretical results are 
well represented by the pseudodiatomic model for effective 
masses of the helium droplet in the desorption reaction of $m_{\mathrm{eff}}/m_{\mathrm{He}}\approx$10
(Rb) and 13 (Cs). Deviations from this model are found for Rb excited to the 
$6p$ state. Photoelectron spectra indicate that the
dopant-droplet interaction induces relaxation into low-lying 
electronic states of the desorbed atoms in the course of the ejection
process. 
\end{abstract}

\maketitle

\section{\label{sec:Intro}Introduction}
Helium nanodroplets are fascinating many-body quantum systems which
feature unique properties such as an extremely low internal 
temperature (0.38\,K), nanoscopic superfluidity, and the ability 
to efficiently cool and aggregate embedded species (dopants). 
Therefore, He nanodroplets are widely used as nearly ideal spectroscopic 
matrices for high resolution spectroscopy of isolated atoms, molecules, 
and clusters~\cite{Stienkemeier:2001,Toennies:2004,Stienkemeier:2006,Barranco:2006}. 

While most studies so far pertain to the structure and time-independent 
spectroscopy of doped He nanodroplets, the dynamics initiated by 
laser-excitation or ionization of either the dopants or the droplets 
themselves moves into the focus of current research. A limited number 
of time-resolved experiments has been carried out with 
pure~\cite{Kornilov:2011,BuenermannJCP:2012} and doped He 
droplets~\cite{Droppelmann:2004,Doeppner:2005,Claas:2006,Przystawik:2008,Mudrich:2009,PentlehnerPRL:2013,PentlehnerPRA:2013,Goede:2013} 
using femtosecond pump-probe techniques. Likewise, theoretical 
models of pure and doped He nanodroplets have mostly been restricted 
to static structure and to excitation
spectrum
calculations~\cite{Barranco:2006,Whaley:1994,Kwon:2000,Chin:1995,Krotscheck:2001}.
Only recently, the development of time-dependent density 
functional theory (TDDFT) methods applicable to microscopic 
superfluids~\cite{Giacomazzi:2003,Lehtovaara:2004}
has opened the way to a time-dependent description 
of doped He droplets in a range of sizes comparable to those used 
in the experiment~\cite{Mateo:2011,Hernando:2012,Mateo:2013,Mateo:2014}.

Dopants consisting of alkali (Ak) metal atoms or molecules are 
particularly interesting due to their weak attractive interaction 
with He droplets which results in their location in shallow dimple 
states at the droplet surface~\cite{Dalfovo:1994,Ancilotto:1995,Stienkemeier:1995}. 
Upon electronic excitation, Ak atoms tend to desorb off the 
He droplet as 
a consequence of the repulsive interaction caused by the overlap
of their extended electronic orbitals with the surrounding 
He~\cite{Reho:2000,Schulz:2001,Callegari:2011}. The only known 
exceptions are Rb and Cs atoms excited to 
their lowest excited states~\cite{Auboeck:2008,Theisen:2011}. 

The dynamics of the desorption process of excited Ak atoms off the 
surface of He droplets has been recently studied in detail
experimentally using the velocity-map imaging technique applied 
to Li, Na and Rb atoms, and theoretically using TDDFT
for Li and Na~\cite{Hernando:2012,Fechner:2012}. The calculated He
droplet response following the dopant excitation process from $ns$ to
$(n+1)s$ states was found to
be quite complex involving different types of density waves propagating 
through the droplet while the Ak dopant is ejected
within a few picoseconds~\cite{Hernando:2012}.
In spite of this, the experiments show that the kinetic energy of the
desorbed atom depends linearly on the excitation energy of the dopant.
This conspicuous result, also reproduced by the TDDFT simulations, 
gives further support to the pseudodiatomic model which has already been 
successfully applied to interpreting the absorption spectra as well as
the ion velocity
distributions~\cite{Stienkemeier:1996,Buenermann:2007,Loginov:2011,Lackner:2011,Fechner:2012}. According to this model, the dynamics of the excited AkHe$_N$ complex 
follows that of a dissociating diatomic molecule~\cite{Busch:1972} 
where He$_N$ plays the role of one single atom in this pseudo-diatom.
The part of the He droplet that effectively interacts with the Ak atom
was found to have an effective mass $m_{\mathrm{eff}}\approx 15$ and
$m_{\mathrm{eff}}\approx 25$ amu  for  Li and Na,
respectively~\cite{Hernando:2012}. 

In the present work, we extend previous ion imaging and TDDFT studies to 
the heaviest stable Ak metal atoms Rb and Cs. We again find linear 
dependences of the ion kinetic energies upon laser photon energy in 
both experiment and theory. From these we infer the effective mass of 
the interacting He droplet for the desorption of Rb and Cs excited to the perturbed $6s$ 
and $7s$ states, respectively. 

While most excited Ak atoms interact repulsively with a He
nanodroplet as a whole, some excited states experience local
attraction with one or a few He atoms. Therefore, as the 
excited Ak atom is expelled from the droplet surface, a bound
AkHe molecule or in some cases small AkHe$_n$, $n=2,3$ 
complexes can
form~\cite{Reho:1997,Droppelmann:2004,Bruehl:2001,Mudrich:2008,Schulz:2001,Giese:2012}. 
These so called `exciplexes' are characterized by having bound vibronic 
states as long as the complex is electronically excited. Upon spontaneous 
decay into the electronic ground state the exciplex decomposes. For such
excited states of the Ak atom, the desorption dynamics may be expected 
to deviate from that described by the simple dissociating
pseudo-diatom model.

In our previous experiment on Rb-doped He droplets excited into 
the $6p\Pi$ state, the ion kinetic energy distributions indicated
that desorption of excited Rb atoms and RbHe exciplexes proceeds 
along the repulsive pseudodiatomic potential which correlates to 
the closest-lying excited $6p$ state of the free Rb atom.
However, the photoelectron spectra clearly revealed that a large 
fraction of the desorbed Rb atoms have electronically relaxed into 
lower-lying levels. The photoelectron spectra contained components 
of the $6p$ state and of lower-lying levels
($4d$ and 5$p_{3/2}$)~\cite{Fechner:2012}.

Previously, droplet-induced relaxation of excited Rb atoms was 
only observed within the 5$p_{3/2,\,1/2}$ fine-structure 
doublet~\cite{Bruehl:2001}. For Rb and Cs injected into bulk 
superfluid He fast relaxation of the lowest excited $p_{3/2}$ 
state into the $p_{1/2}$ and probably to the $s_{1/2}$ ground 
state was found to proceed within $\sim 30$~ps~\cite{Takahashi:1993}. 

For Na-doped He nanodroplets, droplet-induced electronic relaxation was first observed only for higher-lying excitations with principal quantum numbers $n>6$, where the dopant-droplet interaction induces significant mixing of electronic configurations~\cite{Loginov:2011}. In a more recent study, even for the $3d$, $5s$ and $4d$-states the authors found indications for droplet-induced decay into lower-lying levels~\cite{Loginov:2014}. Interestingly, the presence of the relaxation channels was also visible in the speed distributions of the desorbed atoms, which contained multiple components. High-lying Rydberg states were found to completely relax into levels $n\leq 7$. Based on these observations, the authors suggested that droplet-induced relaxation proceeds via level-crossings of the pseudodiatomic potential curves which occur while the local He droplet environment of the excited Na dopant dynamically rearranges. Efficient He droplet-induced electronic relaxation was also observed for barium~\cite{Loginov:2012} and for the transition metal 
atoms silver~\cite{Loginov:2007}, chromium~\cite{Kautsch:2013} 
and copper~\cite{Lindebner:2014}, which are submerged in the droplet interior. 

Note, however, that the light Ak metals Li and Na were not found to electronically relax by droplet interactions when excited into the lowest excited $s$-states (orbital angular momentum $\ell =0$)~\cite{Hernando:2012}. In contrast, in the present study on Rb and Cs atoms in their lowest excited $s$-states we detect \textit{exclusively} relaxed electronic levels in the photoelectron spectra. We discuss the apparent discrepancy between the ion and electron measurements in terms of the desorption dynamics and electron energetics. 

 
\section{Experimental}
The experiments presented here are performed using the same setup 
as described previously~\cite{Fechner:2012}. In short, a continuous beam 
of He nanodroplets with a mean size ranging from 200 to 17000 He atoms 
per droplet is generated by varying the temperature $T_0$ of a 
cryogenic nozzle with a diameter of $5\,\mu$m~\cite{Toennies:2004,Stienkemeier:2006}. 
An adjacent vacuum chamber contains a vapor cell filled with bulk 
metallic Rb or Cs heated to 85$^\circ$C and 70$^\circ$C, respectively. 
In the detector chamber further downstream, the He droplet beam intersects 
a dye laser beam (Sirah Cobra, pulse length $~10$\,ns, pulse 
energy $~10\,\mu$J, repetition rate $1\,$kHz) at right angles in 
the center of a velocity map imaging (VMI) spectrometer. The laser 
is linearly polarized along the direction of the He droplet beam, 
which is perpendicular to the symmetry axis of the VMI spectrometer. 
We record single events per image frame for which the coordinates are 
determined using the centroid method. Velocity-map photoelectron and 
photoion images are transformed into kinetic energy distributions 
using standard Abel inversion
programs~\cite{Vrakking:2001,Garcia:2004}.


\section{Theoretical approach}
In order to model the absorption spectra as well as the dynamic
response of the excited doped He droplets we describe the 
doped He droplets within the Density Functional Theory
(DFT) framework~\cite{Barranco:2006}. The basic ingredients 
of our approach 
are described in detail in Refs. \cite{Hernando:2012,Mateo:2013}. 
Let us just recall that we have used the Born-Oppenheimer approximation
to factorize the electronic and nuclear wavefunctions,
the Franck-Condon approximation which assumes that the
atomic nuclei do
not change their positions or momenta during the electronic
transition, and the 
diatomics-in-molecules approximation (pseudodiatomic model)~\cite{Ellison:1963}.

We have first obtained the structure of the Rb-droplet and Cs-droplet 
complexes in the ground state. Throughout this work we have used the 
Orsay-Trento (OT) density functional~\cite{Dalfovo:1995} neglecting 
the backflow term. The Rb-He and Cs-He ground state pair potentials
$V_X$ have been taken from Ref.~\cite{Patil:1991}.
Due to the large mass of Rb and Cs compared to that of He, we describe 
them as classical particles in the dynamics while their effect in the 
statics is incorporated as an external field acting upon the 
droplet~\cite{Mateo:2013}. Accordingly, the energy of the system is written as
\begin{eqnarray}
E[\rho] &=& \int d \mathbf{r} \,
\frac{\hbar^2}{2m_{\mathrm{He}}}\big|\nabla \sqrt{\rho(\mathbf{r})}\big|^2 +
{\cal E}_{\mathrm{He}}[\rho(\mathbf{r})]
\nonumber \\
&+& \int d \mathbf{r}
\rho(\mathbf{r})V_X(|\mathbf{r}_\mathrm{Ak} -\mathbf{r}|)
\; ,
\label{eq1}
\end{eqnarray}
where
${\cal E}_{\mathrm{He}}$
is the OT potential energy density per unit volume, Ak represents either
the Rb or Cs atom, and $\rho$ is the He particle density. Upon
variation, one obtains the Euler-Lagrange equation that has to be solved
to determine the equilibrium density $\rho_0(\mathbf{r})$ of the 
droplet and the location of the dopant Rb or Cs atom 
$\mathbf{r}_{{\rm Ak}_0}$~\cite{Buenermann:2007}. Schematically,
\begin{equation}
\frac{\delta}{\delta\rho}
\left(
  \frac{\hbar^2}{2m_{\mathrm{He}}}\big|\nabla \sqrt{\rho}\big|^2
  +
  {\cal E}_{\mathrm{He}}
\right)
+ V_X = \mu \; ,
\label{eq2}
\end{equation}
where $\mu$ is the chemical potential of the He droplet that  
throughout this paper is made of $N=1000$ atoms. To explore 
other locations of the Ak atom around its equilibrium position 
in the surface dimple, we have minimized the  energy submitted 
to a constraint~\cite{Mateo:2013}. This will be useful for 
determining the mean kinetic energy of the ejected Ak atom as a 
function of the excess excitation energy.

Equation (\ref{eq2}) has been solved in cartesian coordinates using
a spatial grid of 0.4~\AA{} and a $200 \times 200 \times 250$
points mesh.
The derivatives have been calculated with  13-point formulas.
Extensive use of fast-Fourier techniques has been made to efficiently 
calculate the energy density and dopant-droplet interaction 
potentials~\cite{Hernando:2012,Mateo:2013}.

The dynamics is triggered by the sudden substitution of the Ak-He ground
state pair potential by the excited one. Within 
TDDFT, we represent the He droplet  by a complex effective wavefunction 
$\Psi_{\mathrm{He}}(\mathbf{r},t)$ such that
$\rho(\mathbf{r},t) = |\Psi_{\mathrm{He}}(\mathbf{r},t)|^2$.
The position of the Ak atom  $\mathbf{r}_{\mathrm{Ak}}(t)$
obeys Newton's equation. For excitations involving two $s$ states,
the evolution equations derived in Ref. \cite{Mateo:2013} adopt
a simple form, namely
\begin{eqnarray}
i\hbar\frac{\partial}{\partial t} \Psi_{\mathrm{He}}
&=&
\left[
  -\frac{\hbar^2}{2m_{\mathrm{He}}}\nabla^2 +
  \frac{\delta {\cal E}_{\mathrm{He}}}{\delta \rho(\mathbf{r})}
  +
  V_{ns}(\mathbf{r}- \mathbf{r}_{\mathrm{Ak}})
\right]
\Psi_{\mathrm{He}}
\nonumber
\\
m_{\mathrm{Ak}} \ddot{\mathbf{r}}_{\mathrm{Ak}}
&=&
- \nabla_{\mathbf{r}_{\mathrm{Ak}}}
\left[
  \int d \mathbf{r} \rho(\mathbf{r})
  V_{ns}(\mathbf{r}- \mathbf{r}_{\mathrm{Ak}}) 
\right] \; .
\label{eq3}
\end{eqnarray}
In the above equations, $V_{ns}$ with $n=6(7)$ is the $6s(7s)$ excited
Rb(Cs)-He pair potential~\cite{Pascale:1983}.
The initial configuration to solve
Eqs.~(\ref{eq3}) is  the static dopant-droplet configuration, either
at equilibrium or with the dopant sitting in another position around
the surface dimple, $\Psi(\mathbf{r}, t=0)$=
$\sqrt{\rho_0(\mathbf{r})}$,
$\mathbf{r}_{\rm Ak}(t=0)= \mathbf{r}_{{\rm Ak}_0}$.
The initial velocity of the Ak dopant is set to zero.

Equations~(\ref{eq3}) have been solved using the same grid as 
for the static problem and a time step of  0.5 fs. We have used 
a predictor-corrector method~\cite{Ralston:1960}
fed by a few time steps obtained by a fourth-order Runge-Kutta algorithm.

\begin{figure}
\centering
\includegraphics[width=0.8\textwidth]{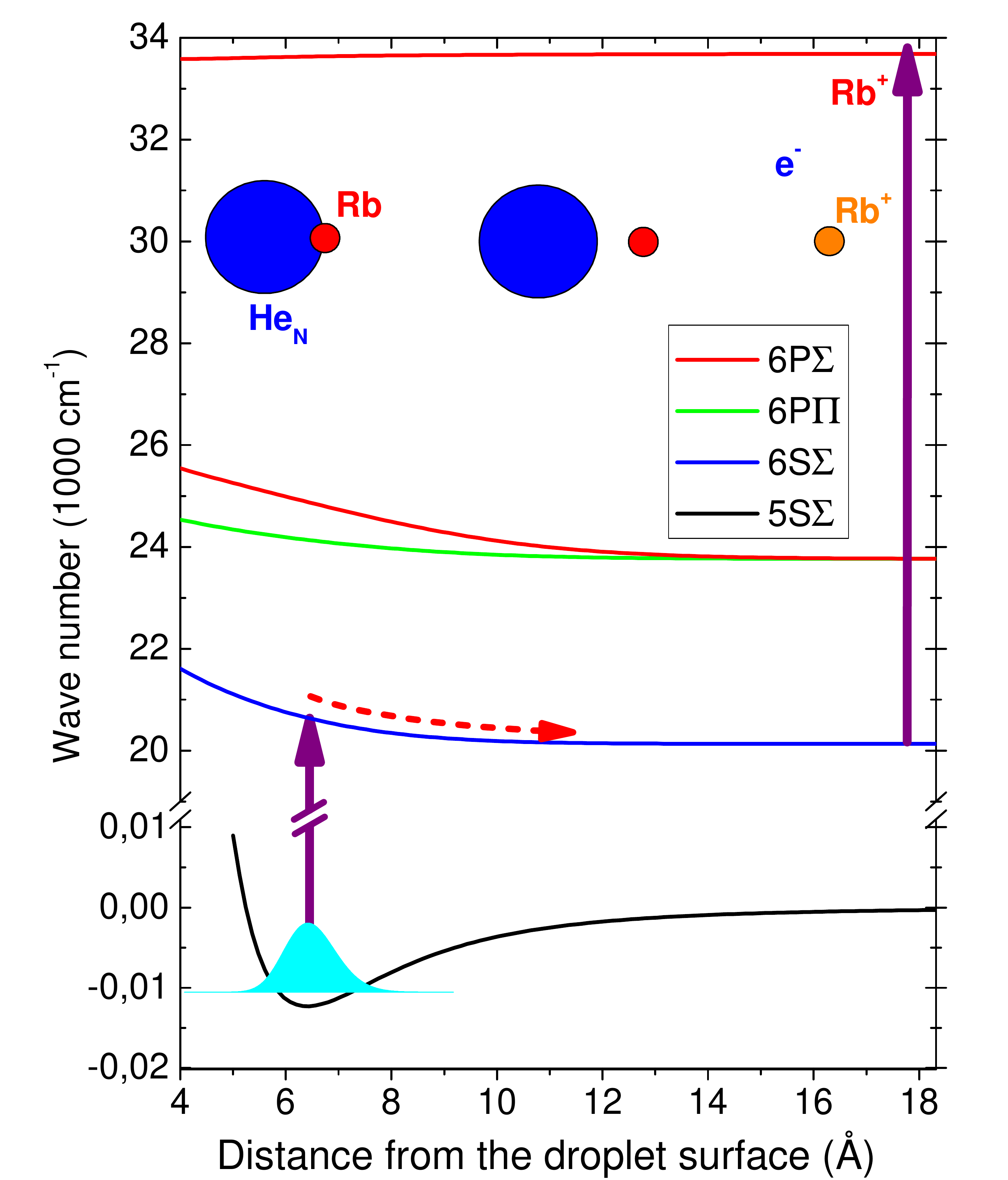}
\caption{Sketch of the excitation and ionization scheme of Rb 
attached to He nanodroplets. Upon excitation of the RbHe$_N$ 
complex to a repulsive pseudodiatomic potential~\cite{Callegari:2011}, 
the Rb atom departs from the droplet surface and is ionized by a second 
photon from the same laser pulse.}
\label{fig:scheme}
\end{figure}

\section{Photoions}
In this work we focus on the $ns\Sigma \rightarrow (n+1)s\Sigma$
transitions of the RbHe$_N$ and CsHe$_N$ pseudo-diatoms, where $n=5,\, 6$ denotes 
the principal quantum number of the atomic ground states of Rb and Cs, respectively. 
The excitation scheme is represented in Fig.~\ref{fig:scheme} for Rb, where we use the pseudodiatomic potential energy curves computed by Callegari and Ancilotto~\cite{Callegari:2011}. 
The ionic potential is obtained by integration of the Rb$^+$-He pair 
potential~\cite{Koutselos:1990} over the He density distribution corresponding 
to the Rb ground state configuration, which we assume to be 
frozen~\cite{Buenermann:2007}. 
Since the Ak-He interaction in the excited $(n+1)s\Sigma$ states is 
purely repulsive, the excited Ak atoms detach from the He droplets 
as neat atoms. Subsequent ionization by the absorption of a second 
photon from the same nanosecond laser pulse yields atomic ions 
which we detect with the VMI spectrometer.

\begin{figure}
\centering
\includegraphics[width=0.8\textwidth]{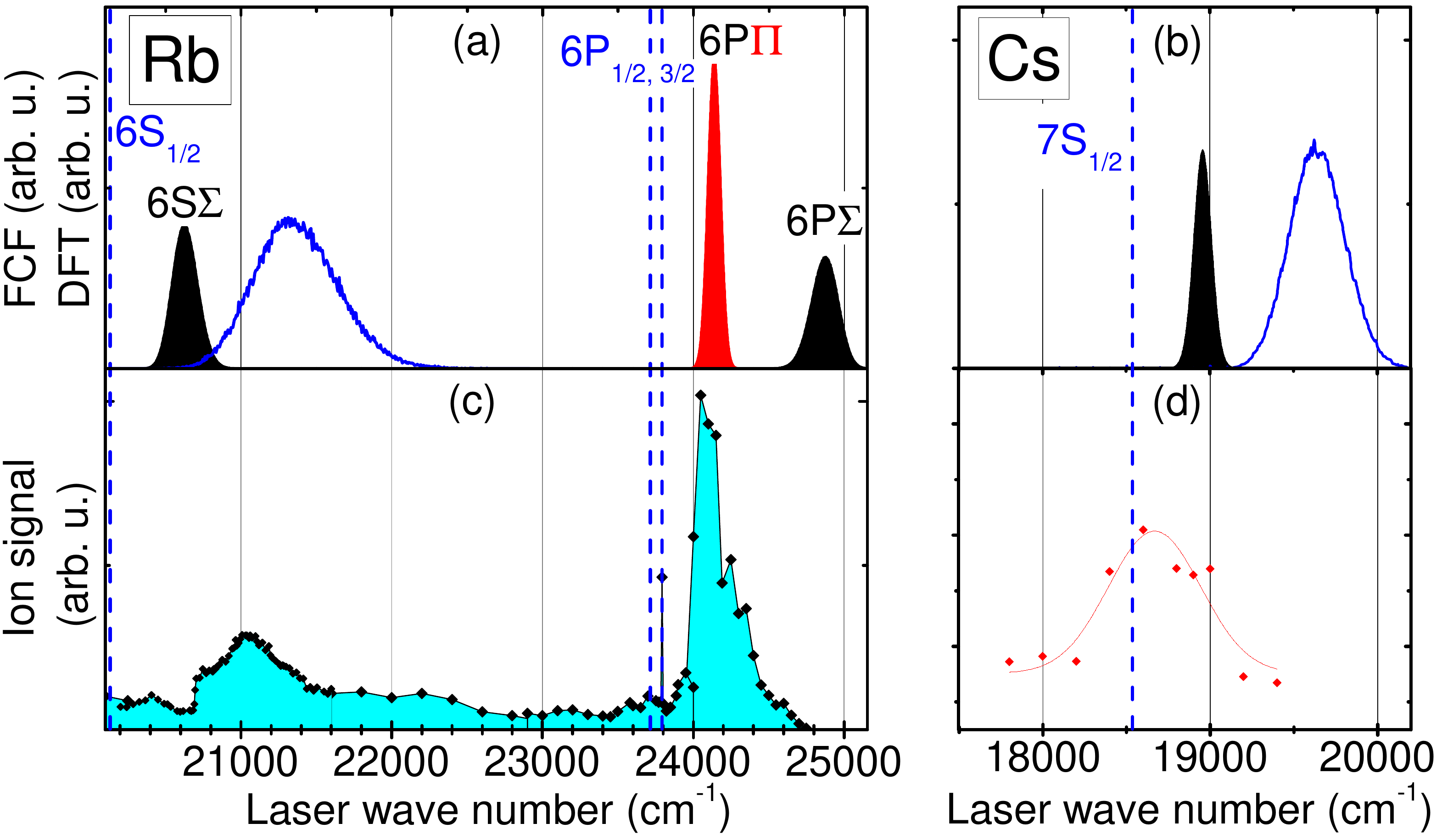}
\caption{Simulated (a, b) and measured (c, d) photoionization spectra of He nanodroplets doped with Rb and Cs. The filled curves in (a, b) show Franck-Condon calculations based 
on Rb-He$_N$ and Cs-He$_N$ pseudodiatomic potentials~\cite{Callegari:2011}; the blue lines 
show the Rb $6s\Sigma$ (a) and Cs $7s\Sigma$ (b) absorption profiles obtained from the present atomic-like DFT sampling method. }
\label{fig:PIspectrum}
\end{figure}

Using these potentials~\cite{Callegari:2011}, we have obtained the Rb and Cs absorption spectra
by calculating wave functions and Franck-Condon factors for the pseudodiatomic transitions using R. LeRoy's program BCONT 2.2 ~\cite{bcont}. The results are depicted in Fig.~\ref{fig:PIspectrum}. We have also calculated the
Rb $5s\Sigma \rightarrow 6s\Sigma$ and Cs $6s\Sigma \rightarrow 7s\Sigma$ absorption band contours by employing the atomic-like DFT sampling method described in Ref.~\cite{Mateo:2011,MateoPRB:2011}, shown in that figure as blue lines. The
vertical dashed lines indicate the atomic transitions. 

Both experimental and theoretical absorption spectra are characterized 
by broad bands which are blue-shifted with respect to the free atomic 
transitions. The blue-shift of the transitions of Rb attached to He 
droplets results from the fact that all excited pseudodiatomic 
potentials are repulsive whereas the ground state is slightly 
attractive (Fig.~\ref{fig:scheme}). The widths of the absorption 
contours reflect the width of the ground state wave function 
which is mapped onto the excited repulsive potential upon 
excitation. While the calculated Franck-Condon profile 
of the Rb $6p\Pi$ transition and the experimental spectrum is
satisfactory, the Franck-Condon profile of the $6s\Sigma$ transition
is slightly red-shifted with respect to the experimental contour,
whereas the DFT result is blue-shifted (Fig.~\ref{fig:PIspectrum} a and c).
The photoionization spectrum of Cs (Fig.~\ref{fig:PIspectrum} d) features a maximum in the range $\sim 18\,300$-$19\,000$ cm$^{-1}$ associated with the 
$7s\Sigma$ transition. The corresponding DFT calculation (Fig.~\ref{fig:PIspectrum} b) yields a peak centered at $19\,635$ cm$^{-1}$, which is again significantly blue-shifted. The DFT calculation thus overestimates the atomic shift, being unclear
which part of the disagreement has to be attributed to deficiencies 
of the model and which to inaccuracies of the excited Ak-He pair
potentials we are using~\cite{Pascale:1983}.

Note that a broad Rb$^+$ ion signal level which features a step around 20\,700~cm$^{-1}$ is measured 
around the $6s\Sigma$ feature, as previously observed in
photoion~\cite{Fechner:2012} and laser-induced fluorescence 
spectra~\cite{Pifrader:2010} around the $6p\Pi$ transition.
This contribution
may be due to photoionization of Rb$_2$ dimers which fragment into Rb$^+$. In particular, at wave numbers below 20\,700 and above 21\,000~cm$^{-1}$ we observe significant Rb$^+_2$ signals in the
tof mass spectrum which could be due to resonance-enhanced ionization via the
$2^3\Sigma_u$ and via the $2^1\Sigma_u$,  $2^1\Pi_u$, or $3^3\Pi_g$ states of Rb$_2$, respectively~\cite{Lozeille:2006}.

\begin{figure}
\centering
\includegraphics[width=0.7\textwidth]{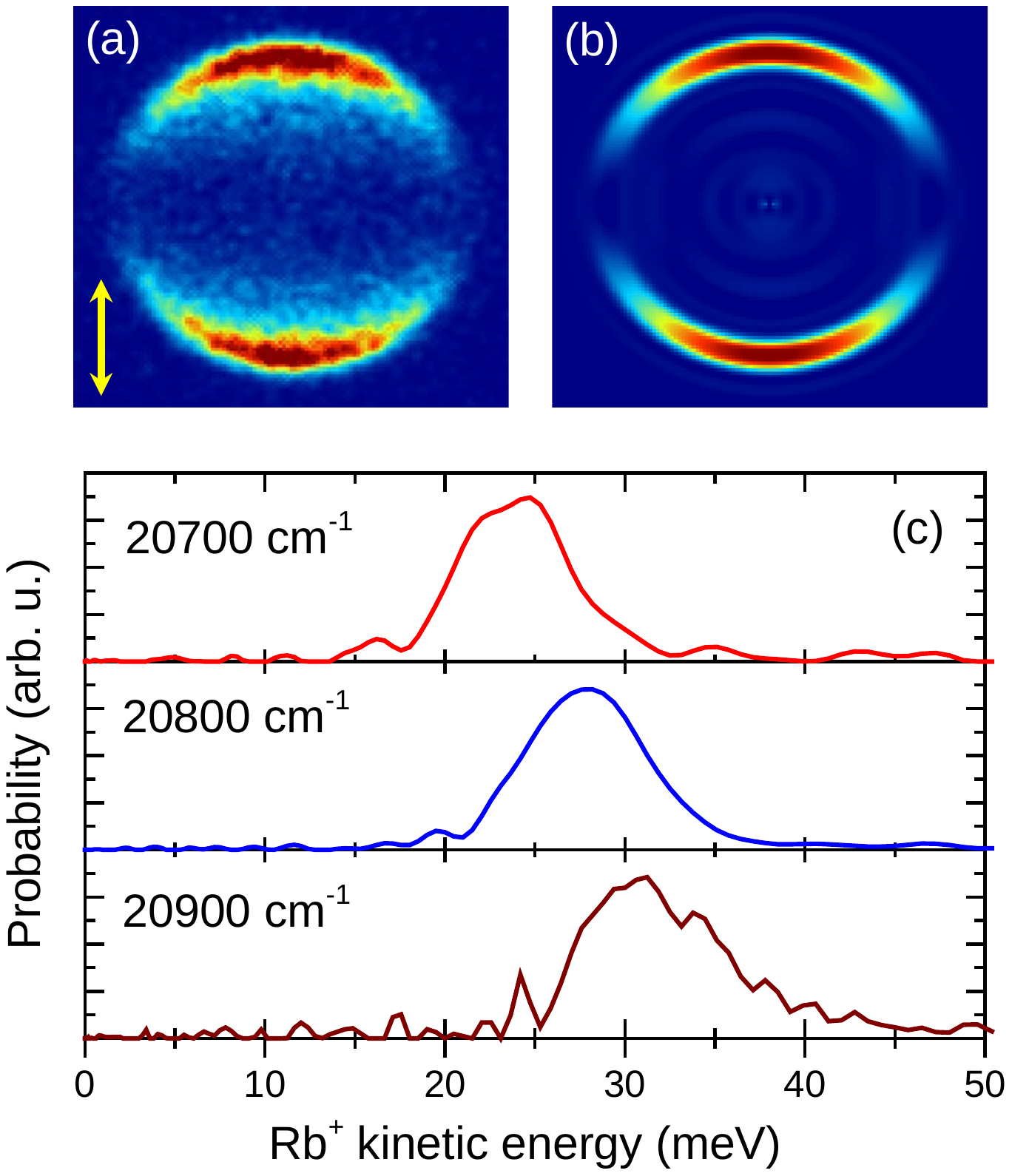}
\caption{Top: Raw (a) and inverse Abel transformed (b) velocity-map
Rb$^+$ ion
images recorded when exciting Rb atoms attached to He nanodroplets on the 
transition $5s\Sigma\rightarrow 6s\Sigma$ at the laser wave number 
$\bar{\nu}=20\,800\,$cm$^{-1}$.
The laser polarization direction is indicated by the vertical arrow. 
Bottom: Rb$^+$ ion kinetic energy distributions inferred from ion images
recorded at the indicated laser wave numbers.
}

\label{fig:ionimages}
\end{figure}

The dynamics of the laser-induced desorption process of Rb and Cs 
atoms is studied by recording velocity-map ion images. 
Fig.~\ref{fig:ionimages} (a) and (b) displays the raw and inverted
Rb$^+$ ion images taken upon excitation to the $6s\Sigma$ state of
the RbHe$_N$ complex at the laser wave number 
$\bar{\nu}=20\,800\,$cm$^{-1}$. The image features a circular 
intensity distribution with a pronounced anisotropy of the angular 
dependence. The intensity maxima are directed along the polarization 
axis of the laser (yellow arrow), as expected for the parallel 
$5s\Sigma\rightarrow 6s\Sigma$
transition~\cite{Hernando:2012,Fechner:2012}.
The velocity-map ion images of Cs recorded at the 
$6s\Sigma\rightarrow 7s\Sigma$ transition around 
$\bar{\nu}=18\,700$ cm$^{-1}$ closely resemble those for Rb. 
In the measurements of Rb excited to the $5p\Pi$ state around
$\bar{\nu}=24\,100$ cm$^{-1}$, the opposite anisotropy is observed 
as expected for a perpendicular $\Sigma\rightarrow\Pi$ transition in
the frame of the pseudodiatomic model~\cite{Fechner:2012}.

From these images we infer the ion kinetic energy distributions (KED) 
by applying inverse Abel transformation and angular integration. 
Typical examples of such KED for excitation around the maximum of 
the $6s\Sigma$ band are depicted in Fig.~\ref{fig:ionimages}.
Similarly to the previous measurements with Li and Na~\cite{Hernando:2012}, 
the KED consist of well-resolved maxima of widths $\sim 70$~meV which shift 
toward higher kinetic energies as the photon energy increases.

\begin{figure}
\centering
\includegraphics[width=0.8\textwidth]{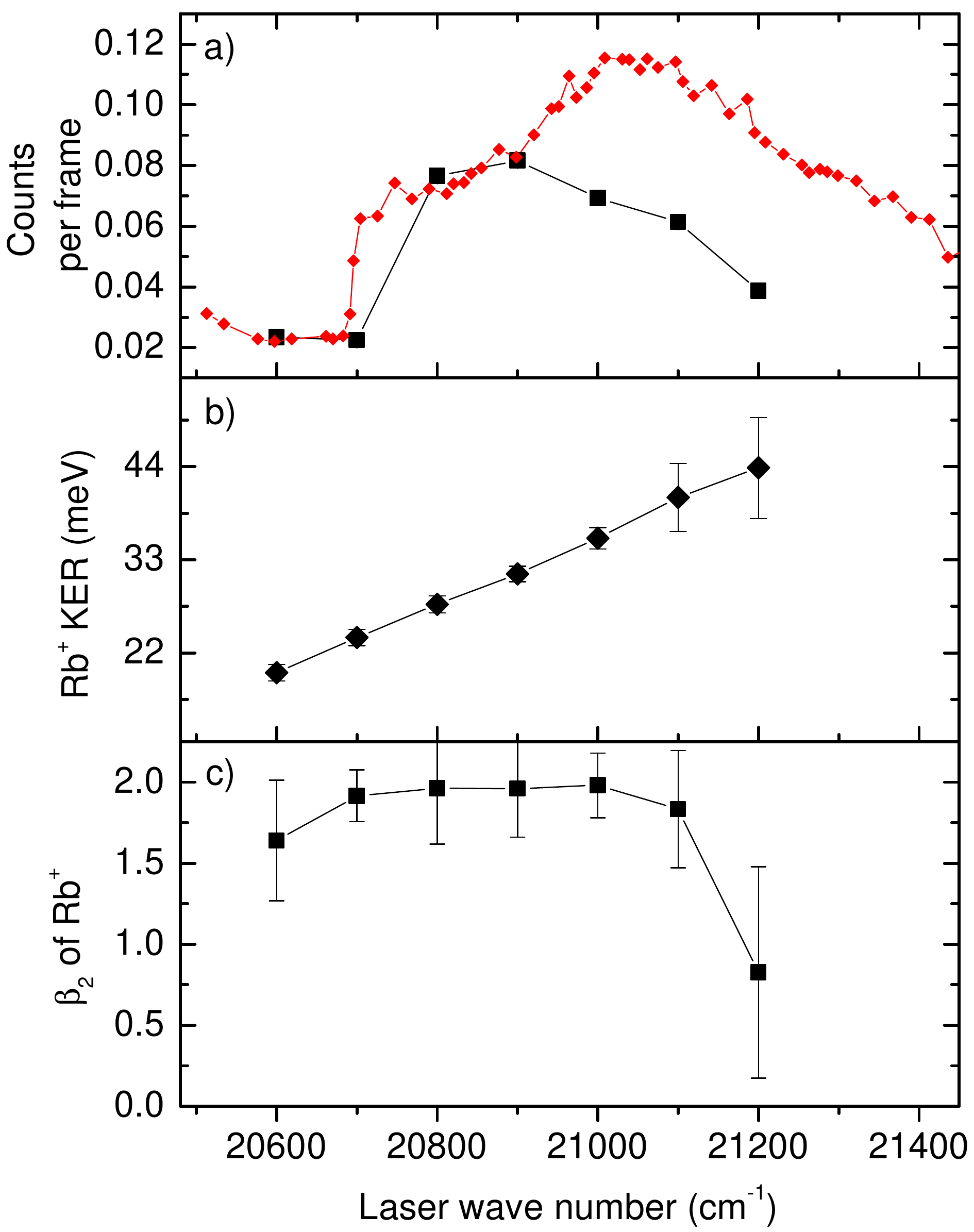}
\caption{(a) Total Rb$^+$ ion counts and tof peak intensity (red), (b) Rb$^+$ mean kinetic energies, (c) anisotropy parameters $\beta_2$ inferred from ion images recorded at various laser wavelengths around the maximum of the Rb $6s\Sigma$ absorption band.}
\label{fig:AnalysisIons}
\end{figure}

Figure~\ref{fig:AnalysisIons} presents a compilation of the results
of the analysis of the ion images recorded around the Rb 
$6s\Sigma$ band as a function of the laser wave number.
The total Rb$^+$ ion counts [black squares in  a)] reproduce
the photoionization spectrum (red diamonds) with some systematic deviation at wave numbers $>$20\,850~cm$^{-1}$ of unknown origin. 
The mean values of the KED inferred from the images, shown 
in Fig.~\ref{fig:AnalysisIons} (b), nearly linearly increases 
with laser wave number. In addition, Fig.~\ref{fig:AnalysisIons} 
(c) shows the variation of the anisotropy parameter $\beta_2$ within a 4 sigma range around the KED intensity maximum inferred from the angular distributions $I(\theta )$ by 
fitting to the general expression for the probability distribution 
of one-photon transitions 
$ I(\theta )\propto 1+\beta_2P_2(\cos\theta )$~\cite{Zare:1972}. 
For laser wave numbers close to the maximum of the $6s\Sigma$ absorption band (20700-21100~cm$^{-1}$) we obtain beta=1.9(3). Within the experimental error this is consistent with the value $\beta_2=2$ expected for excitation of an ideal diatomic molecule at a parallel $\Sigma -\Sigma$ transition. In this case the 
angular distribution of dissociation products takes the form $I(\theta )\propto\cos^2\theta$.

This result nicely confirms the validity of the pseudodiatomic model 
and the assignment of the spectral band to the parallel  
$5s\Sigma \rightarrow 6s\Sigma$ transition. However, in the wings of the absorption peak
we find the anisotropy of the angular ion distribution to be significantly reduced. This may be due to the contribution of fragment ions from Rb$_2$ dimers which are present to a small extent in the droplet beam. Besides, it is conceivable that dynamic deformations of the local He droplet environment during the departure of the Rb atom induce perturbations of the electronic configuration of the excited Rb atom which are not accounted for in the pseudodiatomic picture. 



\begin{figure}
\centering
\includegraphics[width=0.8\textwidth]{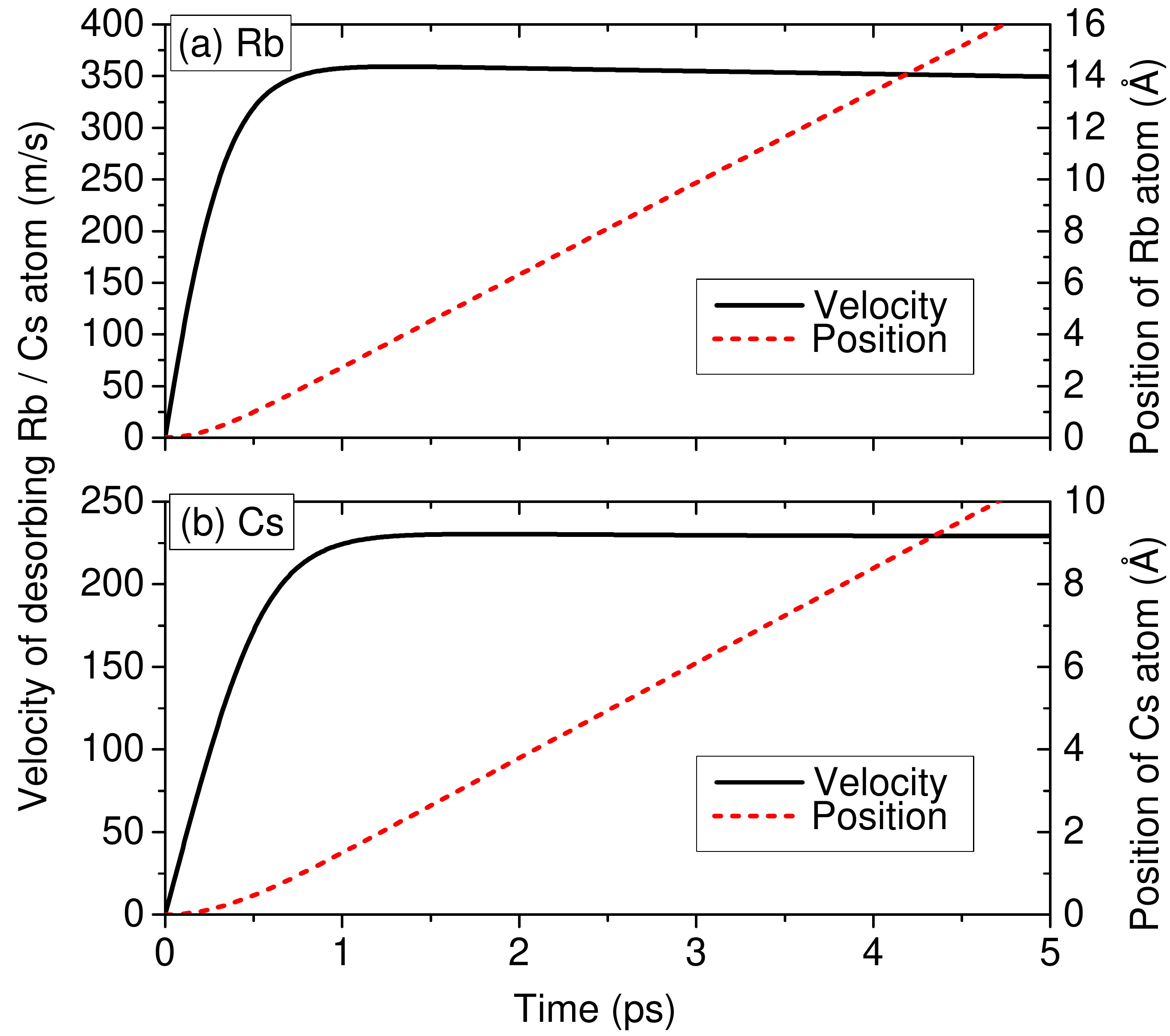}
\caption{(Color online) Velocity (solid line, left scale) and 
displacement from its equilibrium location at the surface 
dimple (dashed line, right scale) of the desorbing Rb (a) 
and Cs (b) atoms as a function of time after excitation of 
the $6s\Sigma$ state.}
\label{fig:VelPosCs}
\end{figure}

In order to obtain more detailed insight into this process, we 
have simulated the ejection of a Rb atom from the nominal $6s$ 
state and of a Cs atom from the nominal $7s$ state using TDDFT 
calculations. 
The velocities and positions as a function of time for Rb 
and Cs ejected from the 
equilibrium position at the surface dimple are shown in 
Fig.~\ref{fig:VelPosCs}. 
It can be seen that the Cs atom reaches an asymptotic
velocity of $\sim 230$~m/s after a time evolution of 	
$\sim 1.25$~ps. By this time, the Cs atom is $\sim 2$~\AA{} 
away from its original 
equilibrium position at the dimple. 
The corresponding values for the Rb atom are 
$\sim 350$~m/s, $\sim 1$~ps and $\sim 2.7$~\AA{}, respectively.
In both cases, the recoil velocity of the He droplet is small,
of the order of 7.5 m/s.
The different evolution of positions and velocities for Rb
and Cs is mainly due to the different masses of the two
species. The fact that both curves are smooth and monotonously 
increasing with time implies that the desorption proceeds impulsively. Thus,
although in the first stages of the dynamics surface vibrations and highly
non-linear density waves are excited in the droplet which take 
a large part of the energy deposited in the system upon photo
excitation, the desorption dynamics is rather insensitive to them.

\begin{figure}
\centering
\includegraphics[width=0.8\textwidth]{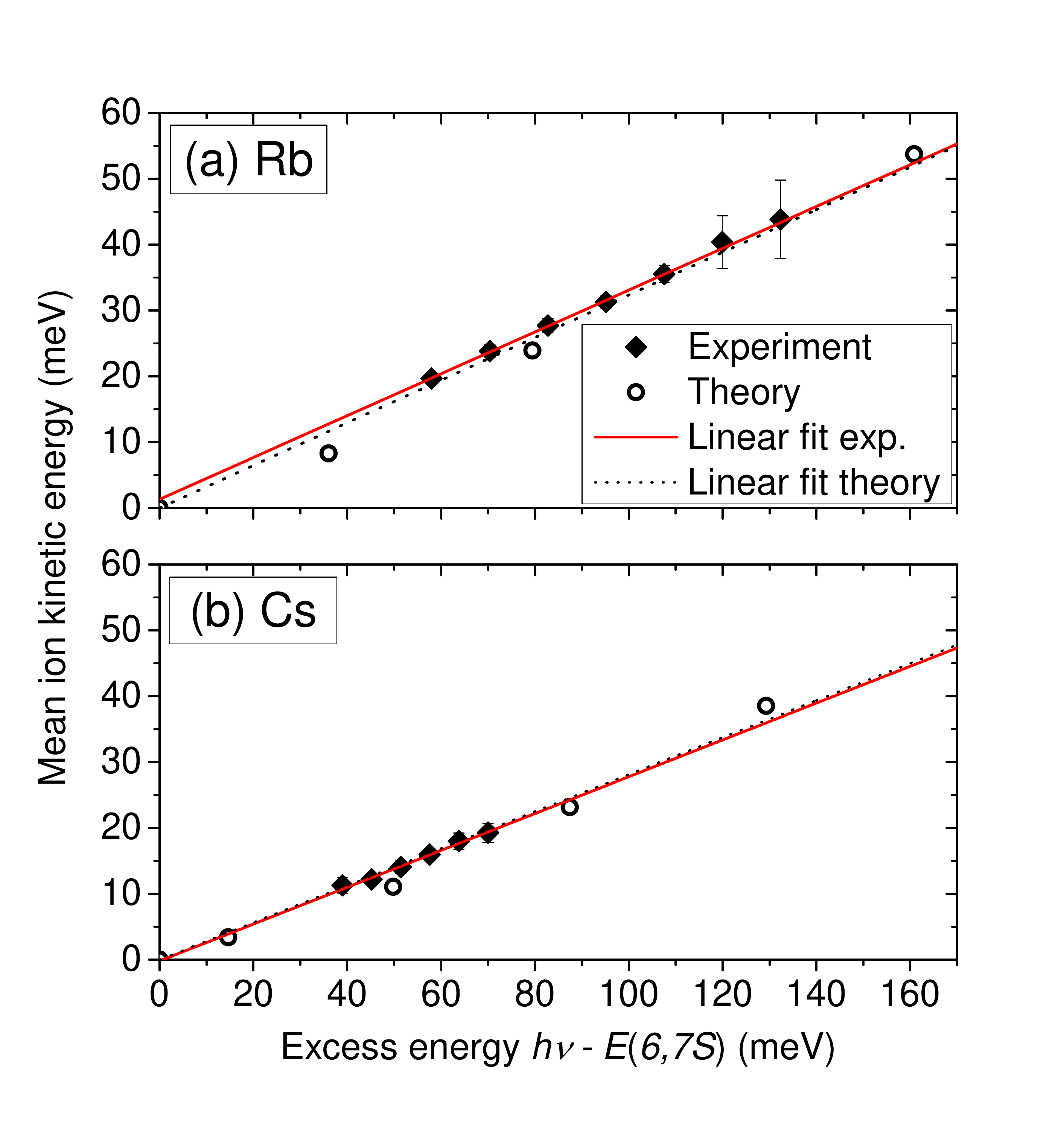}
\caption{Mean kinetic energies of desorbed Rb (a)
and Cs (b) atoms upon excitation to the $6s\Sigma$ and
$7s\Sigma$ states,
respectively. Straight lines: linear fits to the theoretical and
experimental data.}
\label{fig:ExcessEnergies}
\end{figure}

Detailed information about the kinematics of the desorption process can 
be gained from the kinetic energies of the desorbing dopants as a 
function of the excess excitation energy (difference between photon 
energy and internal energy of the free Rb or Cs atom in the $5s$ and 
$6s$ states)~\cite{Hernando:2012}. The results are shown in
Fig.~\ref{fig:ExcessEnergies} and compared to the experimental mean kinetic energies.
The calculated points have been obtained by starting the
dynamic simulation from different positions of the Ak 
obtained by a constrained minimization of the
total energy of the Ak-He$_N$ complex as indicated in Sec. III.

\begin{figure}
\centering
\includegraphics[width=0.8\textwidth]{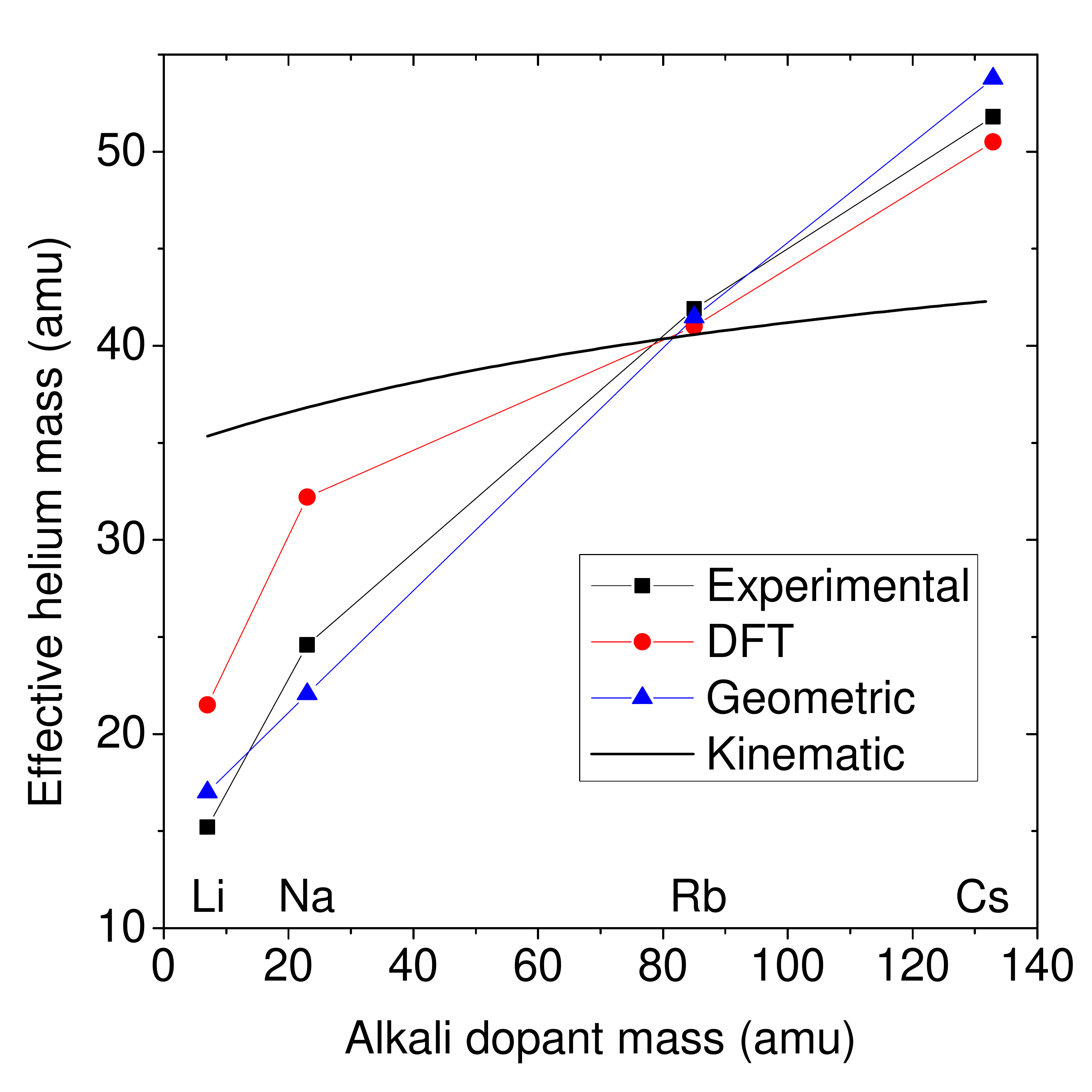}
\caption{Experimental, theoretical and estimated values of the effective mass of the He droplet in the desorption process of various alkali species excited to their first excited $s$-states.}
\label{fig:meff}
\end{figure}

For both Rb and Cs dopants, the kinetic energy displays a linear 
dependence on the excess excitation energy.
This dependence indicates that in spite of its apparent
complexity, the ejection process is well represented by the pseudodiatomic model~\cite{Busch:1972,Hernando:2012}.
Indeed, imposing the energy and linear momentum conservation in the
instantaneous ejection of the Ak atom from the droplet one 
obtains
\begin{equation}
E_{kin} =\eta (\hbar\omega-\hbar\omega_0) \; .
\label{eq4}
\end{equation}
Here, $\omega$ denotes the laser frequency and $\omega_0$ is the atomic transition frequency. Within this model, the value of the slope $\eta$ is related to the mass
$m_{\mathrm{eff}}$ of the part of the He droplet that effectively interacts with the Ak atom~\cite{Hernando:2012} by

\begin{equation}
\eta=  \frac{m_{\mathrm{eff}}}{m_{\mathrm{eff}}+m_{\mathrm{Ak}}}
\Longrightarrow
m_{\mathrm{eff}}=\frac{\eta}{1-\eta}  \, m_{\mathrm{Ak}} \; .
\label{eq5}
\end{equation}
By fitting the experimental and simulation data to the expression Eq.
(\ref{eq4}) one obtains a theoretical value $m_{\mathrm{eff}} \sim 40.7$ amu (10.2 He atoms)
for Rb as compared with the experimental value of $39.6$ amu (9.9 He atoms). 
The corresponding values for Cs are $m_{\mathrm{eff}} \sim 52.0$ amu (13.0 He atoms) from theory and $51.6$~amu (12.9 He atoms) from experiment.
The results for Rb and Cs, together with those obtained for Li and Na in
Ref.~\cite{Hernando:2012}, are collected in Table~\ref{table1} and plotted in Fig.~\ref{fig:meff}. One observes an increase of $m_{\mathrm{eff}}$ with the mass of the Ak atom, as indicated by Eq. (\ref{eq5}), although the prefactor
$\eta/(1-\eta)$ has the opposite behavior. 

The conspicuous dependence of the effective mass of the helium droplet $m_{\mathrm{eff}}$ on the Ak dopant mass $m_{\mathrm{Ak}}$ is expected to be mainly determined by two effects. On the one hand, the geometric structure of the excited Ak-droplet system is different for each species due to slight variations of the ground state equilibrium configuration~\cite{AncilottoZPD:1995} (radius of the surface dimple, distance of the Ak atom from the surface) as well as due to a varying mean radius of the excited Ak atom orbital $r_e$. On the other hand, the kinematics of the dissociation process induces an Ak mass-dependence, irrespective of the differing geometric initial conditions. 

The geometric effect is estimated by computing the geometrical overlap of the electron orbit of the excited Ak atom with the adjacent He atoms of the dimple. Based on the He dimple parameters specified in Ref.~\cite{AncilottoZPD:1995} and on values for the mean orbital radius $\langle r_e\rangle$ we calculate the number of He atoms in the overlap volume $V_{\mathrm{eff}}$ of the excited Ak orbit and He dimple surface, $N_{\mathrm{He, eff}}=V_{\mathrm{eff}}\rho_{\mathrm{eff}}$. Here, $\rho_{\mathrm{eff}}$ is taken as half the bulk value $\rho_{\mathrm{He}}=0.0218$ \AA$^{-3}$ which roughly matches the average He density within the overlap volume due to its location dimple surface where the density smoothly falls off~\cite{AncilottoZPD:1995}. The mean orbital radius is approximated by~\cite{Gallagher:1994,Loginov:2014}
\begin{equation}
\langle r_e\rangle=\frac 3 2 a_0 (n-\delta_l)^2,
\end{equation}
where $a_0$ is the Bohr radius and $\delta_l$ is the quantum defect of the Ak excited state. The corresponding values of $\langle r_e\rangle$ and $N_{\mathrm{He, eff}}$ are added to Table~\ref{table1} and to Fig.~\ref{fig:meff}. 

The kinematic effect of the varying mass of the desorbing Ak atom is probed by solving the classical equations of motion of the Ak atom being repelled off a linear chain of effective, mutually non-interacting He atoms, each containing the mass of 7 He atoms which roughly equals the number of He atoms in the first surface layer next to the Ak dopant~\cite{AncilottoZPD:1995}. The initial spacing between the He ``layers´´ is taken as the average distance between He atoms in the droplets, $3.6$ \AA~\cite{Peterka:2007}. The distance between the Ak atom and the fist He ``layer´´ is held fixed at $5.5$ \AA~and the same Ak-He interaction potential $V_{\mathrm{Ak-He}}(d)=0.2\exp\left(-d/3-1\right)$ (in atomic units) is used for all Ak species. The trajectories of the Ak atoms closely follow those shown in Fig.~\ref{fig:VelPosCs} and the trajectories of the He layers show that mostly the first He layer participates in the desorption dynamics. Accordingly, the effective mass of the He droplet (approximated by the linear chain of atoms with the mass of the He layers) only slightly exceeds the mass of the first He layer, 28~amu, see the solid line in Fig.~\ref{fig:meff}. 

While the He effective mass in this simple kinematic model matches the experimental and DFT values for Rb, the variation as a function of Ak dopant mass is not sufficiently well reproduced (Fig.~\ref{fig:meff}). The simple estimate based on the geometric Ak-He orbital overlap, however, shows a strong variation of the effective mass in surprisingly good agreement with the experimental values. We therefore conclude that the difference in the number of interacting He atoms for the different Ak species is likely related to the difference in the dimple structure and excited electron orbit rather than to the kinematics of the desorption process.

\begin{figure}
\centering
\includegraphics[width=0.6\textwidth]{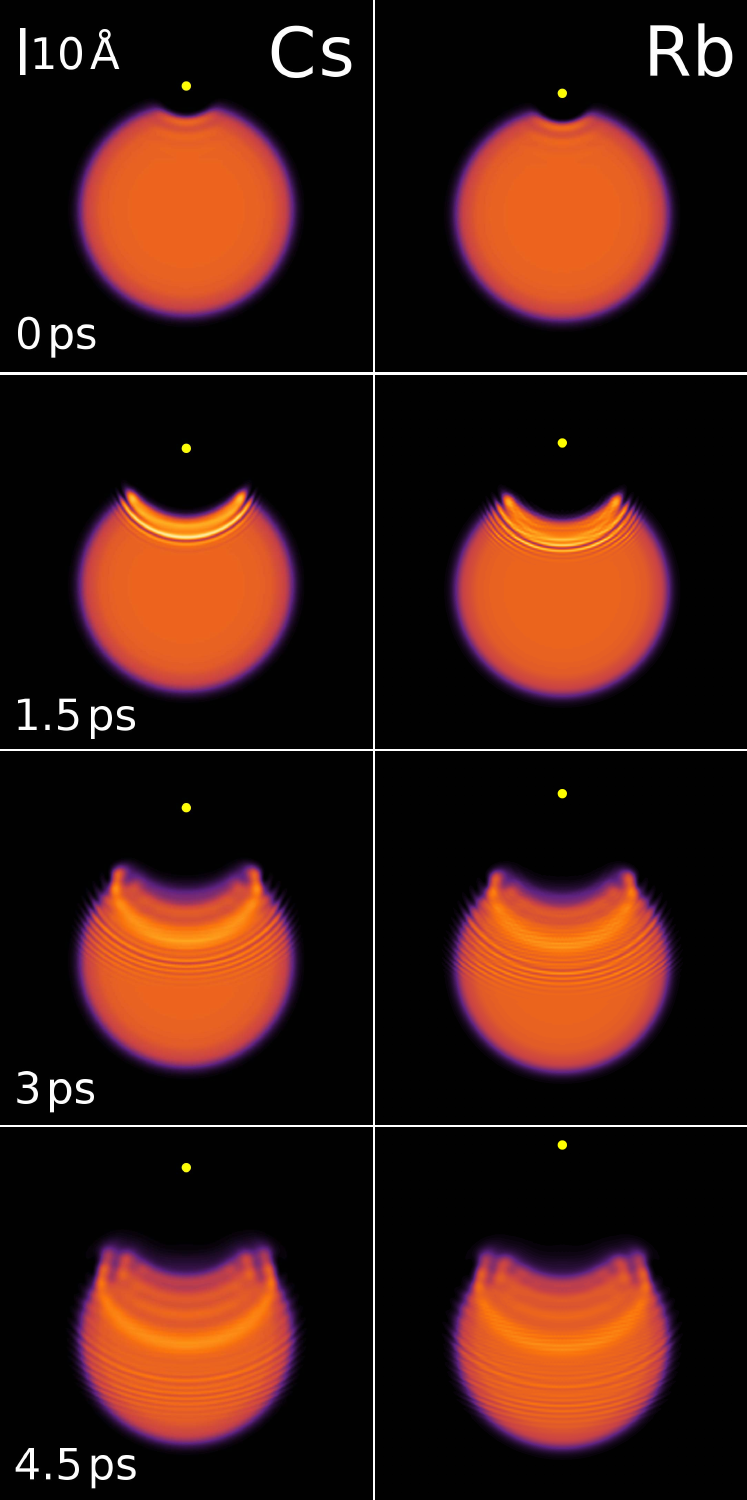}
\caption{Evolution of the He density distributions 
of the CsHe$_{1000}$ (left column) and RbHe$_{1000}$ (right column) systems
after excitation to their $(n+1)s\Sigma$ states. (Multimedia view)}
\label{fig:DensityPlots}
\end{figure}

\begin{figure}
\centering
\includegraphics[width=0.7\textwidth]{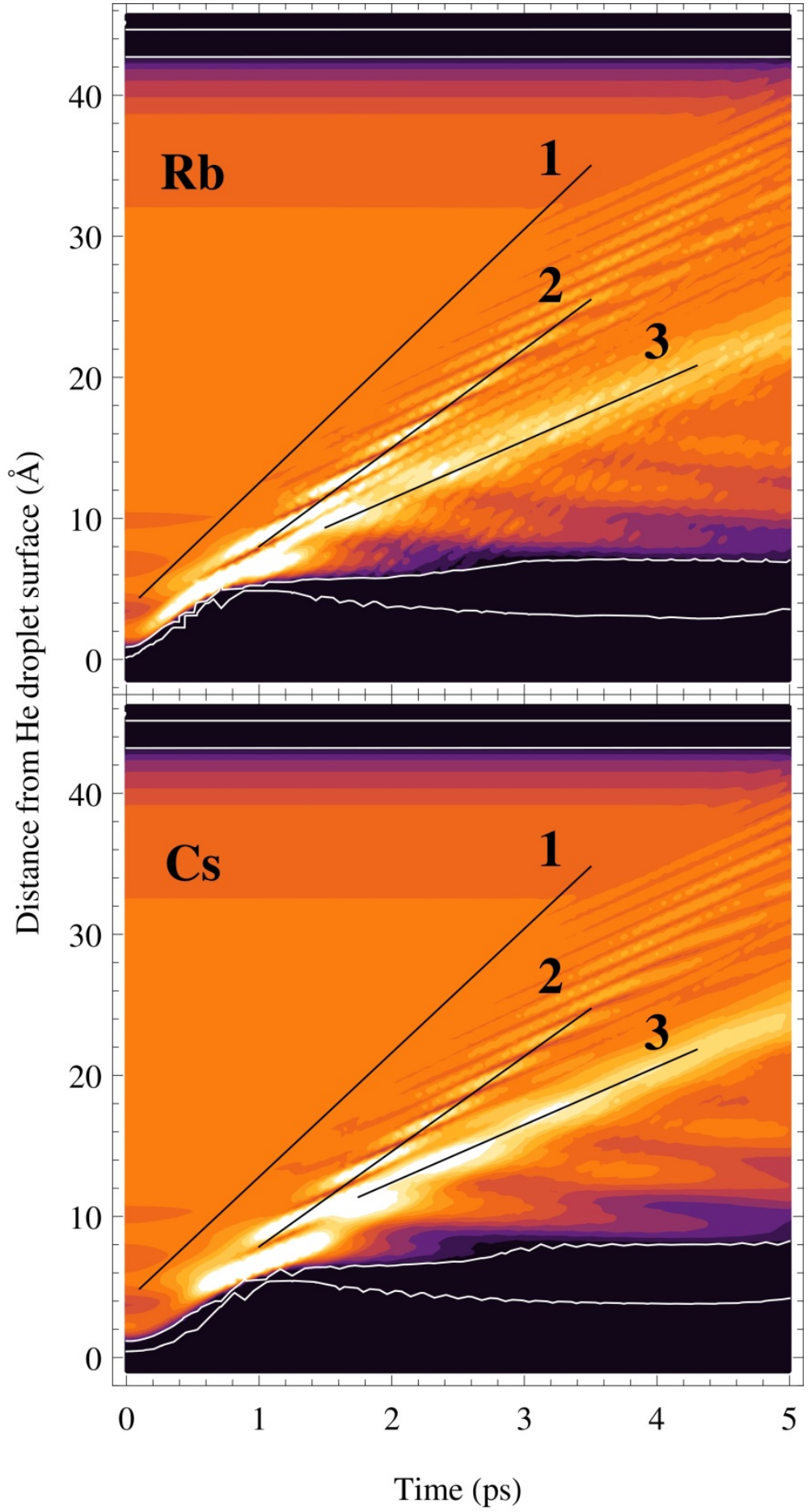}
\caption{Evolution 
of the He density profile of the AkHe$_{1000}$ system along the symmetry axis. Three supersonic wave fronts are identified and labeled by 1 to 3. Equidensity lines corresponding to 0.5 and 0.1 times the He saturation density, 0.0218 \AA$^{-3}$, representing
the surface region of the droplet, are shown in white.
}
\label{fig:waves}
\end{figure}

The detailed picture of the dynamics of the He droplet upon excitation of
the Ak atom is obtained from the DFT calculations. Fig.~\ref{fig:DensityPlots} shows the evolution of the CsHe$_{1000}$ and RbHe$_{1000}$ complexes after the Ak atom has been excited.
It can be seen the dramatic changes in the droplet density caused by the
excitation and subsequent ejection of the dopant.

Figure~\ref{fig:waves} shows the evolution of the He cross-sectional 
density profiles of a He$_{1000}$ droplet doped with a Rb and a Cs 
atom for the first 5 ps. Initially, the droplet extends along the $z$
symmetry axis
from about 0 to 44~\AA{}, and the Ak atom is located in a dimple at the
droplet surface (near $z=0$). Excitation of the Ak atom to the $(n+1)s$
state causes
the dimple first to deepen due to the highly repulsive  
Ak-He interaction in the $(n+1)s\Sigma$ state. The associated
compression
of the He droplet lasts up to $\sim~1$~ps, as shown in the figure.
Following
this compression, the He surface bounces back and the dimple starts 
being filled. The more distant part of the droplet (near $z=42$~\AA{})
is unperturbed and at rest, indicating that during these first ps the
energy deposited in
the droplet goes to its internal excitation 
and not to its center-of-mass motion.

Figures~\ref{fig:DensityPlots} and~\ref{fig:waves} reveal that the 
excitation of the Cs and Rb atoms launches highly non-linear density waves into the droplet. 
In the case of Rb, the first perturbation front, labeled as 1, moves at
$\sim 900$ m/s. This perturbation generates carrier waves with a phase velocity
of $\sim 430$ m/s, modulated by supersonic envelope fronts with growing
intensity. The ones with highest intensity, labeled as 2, have a group
velocity of $\sim 700$ m/s. Next, a high intensity wave appears traveling at
$\sim 410$ m/s (labeled as 3), which generates secondary waves propagating
backwards. 
In the case of Cs, the velocities of the fronts are
880 m/s, 675 m/s, and 410 m/s, respectively.
A similar behavior was found in 
Ref.~\cite{Hernando:2012} for Na and Li atoms.

\begin{figure}
\centering
\includegraphics[width=0.8\textwidth]{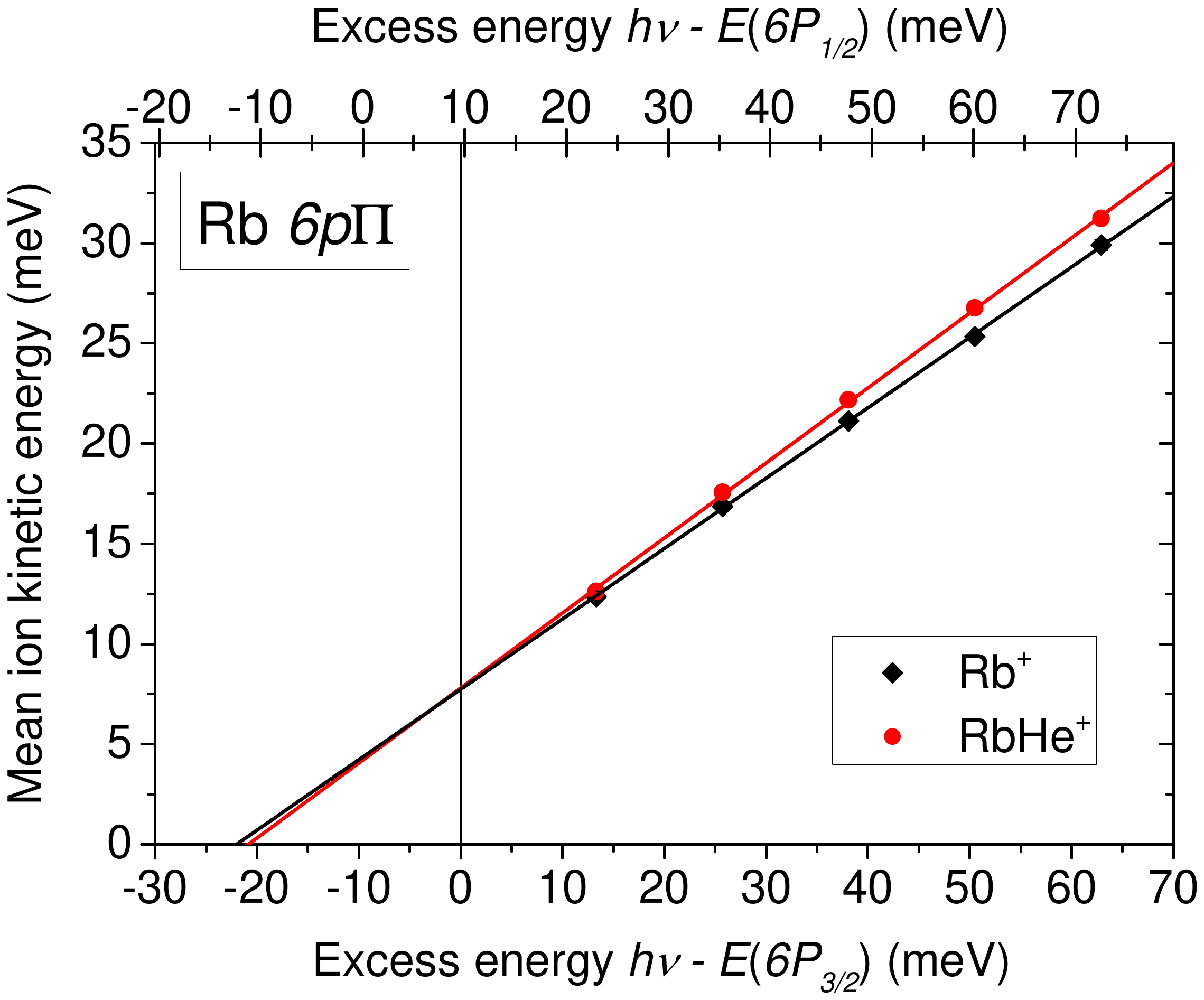}
\caption{Experimental mean kinetic energies of Rb atoms and of RbHe 
exciplexes ejected out of He droplets upon excitation of the 
$6p\Pi$ state of the RbHe$_N$ complex. The lines are linear fits to the
data.} 
\label{fig:ExcessEnergiesRb6pPi}
\end{figure}

As an extension of our previous ion imaging measurements at 
the Rb $6p\Pi$ band~\cite{Fechner:2012} we analyze here the
mean ion kinetic energy as a function of the excess
energy. Since in the 
$6p\Pi$ configuration the Rb-He pair potential along the internuclear axis is
attractive~\cite{Pascale:1983}, RbHe exciplexes are 
formed with roughly 40\% relative abundance~\cite{Fechner:2012}. 
Therefore, we record ion images for Rb$^+$ and RbHe$^+$ ions 
separately and extract the mean ion kinetic energies for each 
of the two species. 

Figure~\ref{fig:ExcessEnergiesRb6pPi} shows that the data points lie on
a straight line that surprisingly intercepts the abscissa at
a finite value of the excess energy of about $-22$ meV. Using 
Eq. (\ref{eq5}), from the slope of the line $\eta$ one obtains 
$m_{\mathrm{eff}}=46.0$ (11.5 He atoms) for Rb, and
$m_{\mathrm{eff}}=53.2$ amu (13.3 He atoms) for RbHe, slightly larger than the
corresponding value for the $6s\Sigma$ state.

The fact that the extrapolation of the $6p\Pi$ experimental data to
zero kinetic energy yields a finite energy shift at zero
kinetic energy, at variance with the extrapolation of the $6s\Sigma$
data, discloses an intrinsic limitation of the method used to
analyze the results. The pseudodiatomic approximation, even if appropriate
for the description of a direct dissociation via a
purely repulsive state, does not account for other effects which are
present in the dissociation kinematics of the $(n+1)p$ excitation.
In the case of the $6p\Pi$ state of Rb, the dopant-He interaction contains both
repulsive and attractive contributions, the latter inducing the formation of
exciplexes. It is conceivable that the binding energy of the RbHe
exciplex may be converted into additional translational energy upon desorption of RbHe. This interpretation has recently been invoked to rationalize the negative excess energy offset measured for NaHe exciplexes formed upon excitation of Na into the droplet-perturbed $3d$ state~\cite{Loginov:2014}. The binding energy of RbHe in the $6p\Pi$ state amounts to about 8~meV, which does not account for the observed energy shift alone.
Additional internal energy may be released into translational motion of the desorbing Rb by
droplet-induced relaxation of population from the upper $6p_{3/2}$ into the lower $6p_{1/2}$
spin-orbit state of Rb. In that case, the excess energy axis would be 
down-shifted as represented by the horizontal top scale of Fig.~\ref{fig:ExcessEnergiesRb6pPi} provided
the droplet effective mass is the same ($m_{\mathrm{eff}}=46.0$ amu) for this additional acceleration of the Rb atom due to spin-orbit relaxation. However, the 
atomic spin-orbit splitting (9.6~meV) does not fully account for the observed shift. Only the assumption that both spin-orbit and binding energy of the RbHe exciplex are fully converted into translational energy would explain the energy offset for RbHe. The nearly coinciding kinetic energies of Rb and RbHe may indicate that Rb$^+$ ions are actually produced by dissociative ionization of RbHe, the latter being the dominant product of the desorption reaction.

Thus, it seems that the pseudodiatomic model no longer strictly applies when the internal degrees of freedom of the constituent atom are involved in the dynamics.
Note that for the case of the desorption of sodium (Na) atoms excited to the 3$p$ state 
deviations from the pseudodiatomic model were also observed~\cite{LoginovPhD:2008}. However, in contrast to the Rb case discussed here, a positive value for the abscissa intercept was found. TDDFT studies of Ak atoms ejected from the $(n+1)p$ excited states could help elucidate this open issue, but improved Ak-He pair potentials have to be previously obtained.

\section{Photoelectrons}
\begin{figure}
\centering
\includegraphics[width=0.7\textwidth]{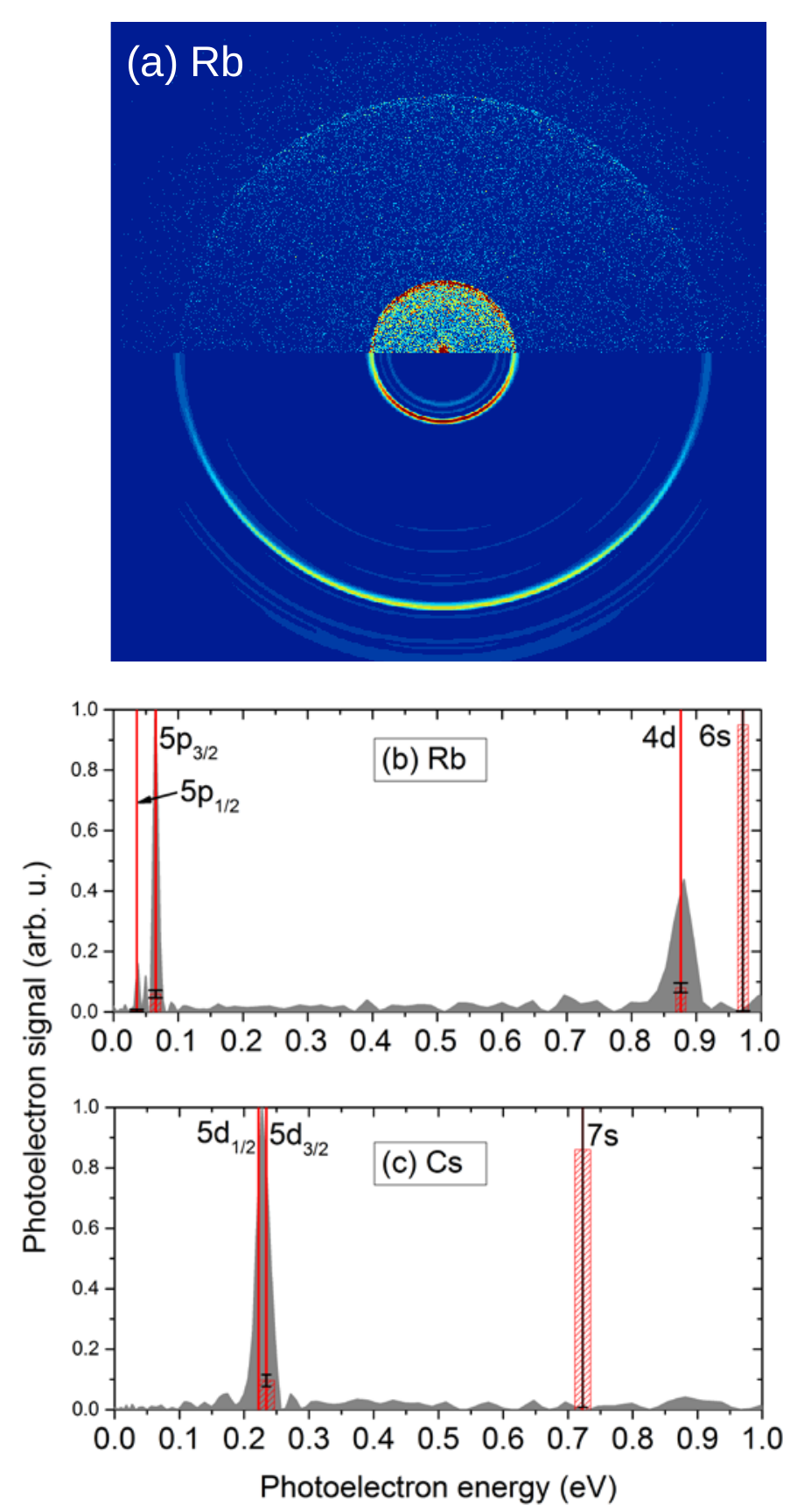}
\caption{(a) Raw (upper half) and Abel inverted photoelectron image 
(lower half) of Rb$^+$ ions recorded with Rb-doped He nanodroplets 
excited into the $6s\Sigma$ state at the laser wave number
$\bar{\nu}=21\,400$~cm$^{-1}$. (b) and (c) Photoelectron spectra of 
Rb and Cs inferred from images recorded at the laser wave numbers 
21\,400~cm$^{-1}$ and 18\,700~cm$^{-1}$, respectively. The vertical
bars represent the relative populations in the respective atomic states.}
\label{fig:PES}
\end{figure}

Complementary information about the dynamics following laser 
excitation of Ak atoms attached to He nanodroplets is obtained from 
imaging photoelectrons. In the experiment, velocity-map photoelectron
images are obtained by simply reversing the polarity of the voltages 
applied to the repeller and extractor electrodes~\cite{Fechner:2012}. 
A typical raw and inverse Abel transformed image recorded at the laser 
wave number $\bar{\nu}=21\,400$~cm$^{-1}$ is depicted in the upper and 
lower half of Fig.~\ref{fig:PES} (a), respectively. The image clearly 
contains three separated ring structures, indicating that ionization 
occurs out of three Rb atomic orbitals. The faint ring structure between the two inner rings in the low half of Fig.~\ref{fig:PES} (a) is an artifact of the inverse Abel tranformation caused by the limited statistics. As for the ion images, 
we again convert the electron images into angular distributions and 
electron kinetic energy spectra. The latter are shown in Fig.~\ref{fig:PES} 
(b) for Rb and in (c) for Cs.

Surprisingly, all the photoelectron spectra recorded within the Rb
$6s\Sigma$ band reveal contributions of the Rb $5p_{1/2}$, $5p_{3/2}$
and $4d$ atomic levels. No electron signal associated with the $6s$ state is detected within the noise level, although the $6s$ state is the dominant atomic
component of the originally excited $6s\Sigma$ state of the RbHe$_N$
complex~\cite{Callegari:2011}. The same holds for Cs excited 
to the $7s\Sigma$ state. Only one peak is present in the spectrum
due to ionization out of the Cs $5d$ state. The $6p$ states,
which are probably populated as in the Rb case, are not 
detected because of insufficient photon energy for one-photon ionization 
of the $6p$ states.

This result is at odds with the previous measurements of Li and Na excited to $(n+1)s$ states and to our previous photoelectron spectra recorded at 
the Rb $6p\Pi$ band where on resonance the dominant photoelectron signals came from
the correlating atomic $6p$ state. The lower lying $4d$ and $5p_{3/2,\,1/2}$ 
states became particularly apparent for off-resonant excitation. 
In the present case, however, the absence of the Rb $6s$ and Cs $7s$ photoelectron signals is probably due to the particularly small 
photoionization cross sections of about $0.01$ Mb which result from Cooper minima close to the laser wave numbers used in the experiment~\cite{Moskvin:1963,Lahiri:1986}. For comparison, the detected states have photoionization cross sections $>10$ Mb~\cite{Aymar:1984,Lahiri:1986}. 



In the case of Na attached to He nanodroplets, the appearance of 
lower-lying atomic states was attributed to the short radiative life
time of the excited level as compared to the laser pulse length~\cite{Loginov:2011}. In 
the Rb and Cs cases, however, as for the Rb $6p\Pi$ state previously
studied~\cite{Fechner:2012}, the lifetimes of the free Rb and Cs
atoms in the $6s$ and $7s$ states ($\sim 50$~ns~\cite{Heavens:1961,Lahiri:1986}) by far
exceed the laser pulse length (9~ns). Moreover, the appearance of the $4d$ state of Rb and of the $5d$ state of Cs cannot be explained by spontaneous radiative decay due to selection rules. Merely the Rb $5p$ photoelectron signal may contain a contribution from radiative decay. Therefore, we attribute the population of lower-lying electronic states to He droplet-induced relaxation. Whether this relaxation mechanism is predominantly non-radiative or whether the dopant-droplet interaction induces fast radiative decay even at nominally forbidden transitions cannot be determined from theses measurements.

The vertical bars in Fig.~\ref{fig:PES} (b) and (c) depict the relative populations of the detected states as inferred from the peak integrals weighted by the photoionization cross sections. The corresponding values of the undetected Rb $6s$ and Cs $7s$-states reflect the noise level and can only be considered as upper bounds. Thus, while the populations of the Rb $6s$ and Cs $7s$ states and of the lower lying states (Rb $5s$ and Cs $6s$, $6p$) are undetermined, the Rb $5p_{3/2}$ and $4d$ states are nearly equally populated. When assuming that the Rb $5p$ level is populated purely by radiative decay, this population corresponds to a fraction of about 13\% of the original $6s$ population whereas the $5p_{1/2}$ state is populated only by 7\%~\cite{Heavens:1961}. However, the fact that the measured population of the $5p_{1/2}$ state only amounts to about 8\% of the $5p_{3/2}$ population indicates that an additional droplet-induced decay process is active. In the case of Cs only the $5d$ state is detected so no quantitative comparison with other states can be made.

The detection of photoelectrons exclusively out of relaxed states seems to contradict the results from ion imaging which clearly demonstrate that desorption proceeds according 
to the pseudodiatomic model for a fixed $(n+1)s\Sigma$ electronic configuration. For Na excited into the droplet-perturbed states $5s$ and $4d$, the presence of relaxation channels was also observed in the speed distributions of the desorbed atoms~\cite{Loginov:2014}. A broad, nearly laser wavelength independent component extending out to velocities $\sim$1500 m/s (kinetic energy $\sim$270 meV) was assigned to atoms having undergone relaxation to the lower $3d$-level. However, in the present experiments on Rb and Cs in the $(n+1)s$-state, no such broad component of the ion distributions is observed (see Fig.~\ref{fig:ionimages}). The range of kinetic energies observed  (Fig.~\ref{fig:ExcessEnergies}) matches well the values expected for dissociation to proceed along the Rb and Cs $(n+1)s\Sigma$ potentials, see Fig.~\ref{fig:scheme}. 

Furthermore, we have considered the possibility that the photoelectron peaks from relaxed states could be associated with Rb$_2^+$ and Cs$_2^+$ dimer ions. However, the relative yield of dimers falls far below the proportion of photoelectrons in relaxed states. Besides, the dependence of the signal intensity of the relaxed electrons on the Rb and Cs vapor pressure in the doping cells clearly indicates that these electrons correlate to Rb$^+$ and Cs$^+$ atomic ions. In addition, the possible correlation of these electrons with large ion masses, resulting from unfragmented ion-doped He droplets, was probed by performing dedicated time-of-flight measurements using a different detection unit. The measured proportion of large cluster ions to Rb$^+$ again stayed well behind that of relaxed electrons to (undetected) electrons out of the Rb $6s$-state. However, due to the uncertainty in determining the relative detection efficiency for large ions, this possibility cannot strictly be ruled out. 




Thus, our observations seem to imply that electronic relaxation 
occurs with some time delay with respect to the strong repulsive interaction
which accelerates the dopant atom away from the droplet surface. He induced electronic couplings may be facilitated by the formation of a compressed shell of He atoms around the dopant in the course of desorption ($t=$0.5-1.5~ps, see Fig.~\ref{fig:DensityPlots}). 
Theoretical modeling of the coupled electron dynamics of excited dopant-droplet complexes 
as well as time and mass-resolved ion and electron imaging experiments are 
needed to resolve this puzzling issue. 

\begin{figure}
\centering
\includegraphics[width=0.8\textwidth]{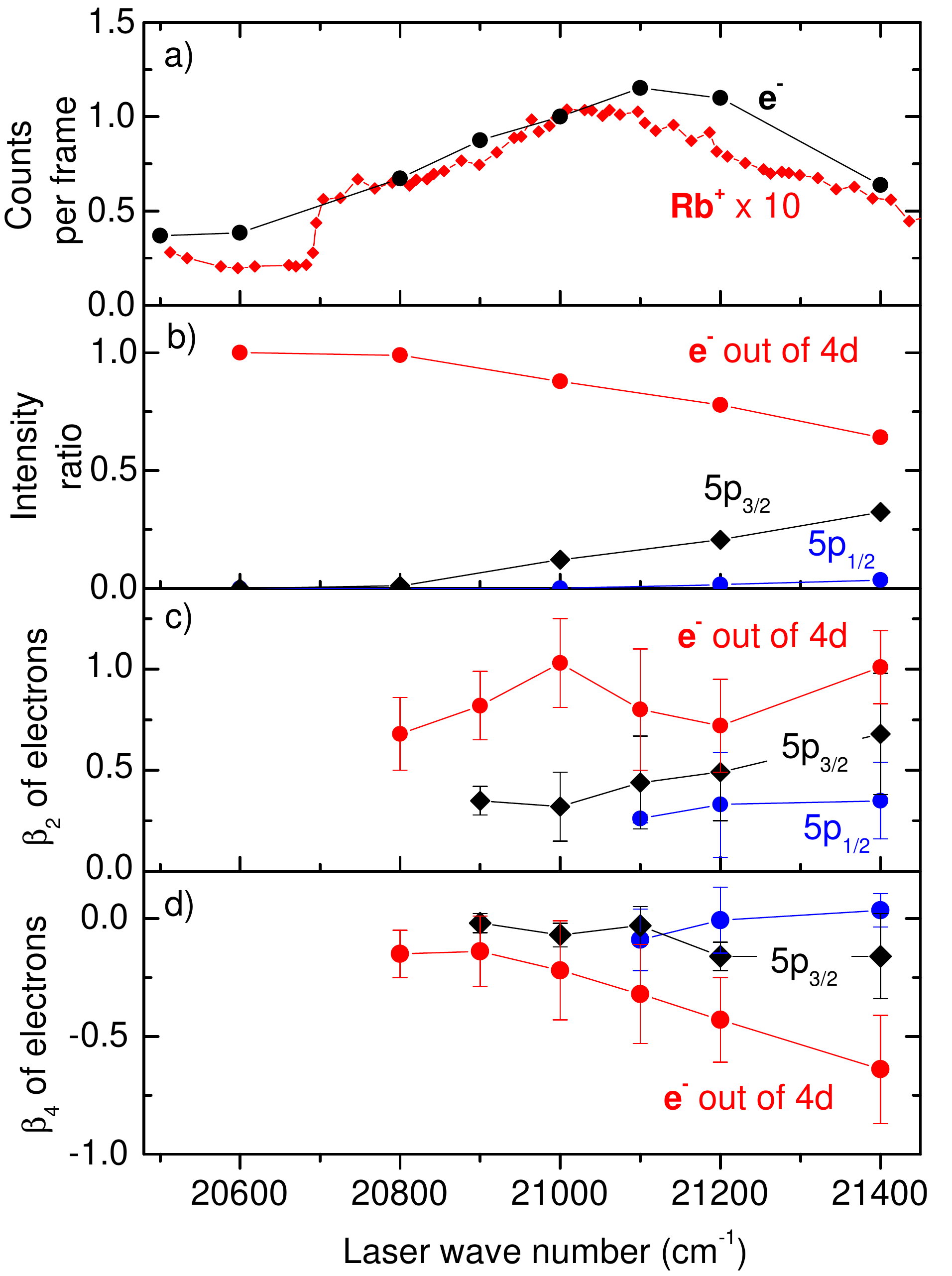}
\caption{Total photoion signal and photoelectron counts (a), relative 
abundances of electrons out of different atomic states (b), and 
anisotropy parameters $\beta_2$ (c) and $\beta_4$ (d) inferred 
from electron images recorded a various laser wavelengths 
around the maximum of the Rb $6s\Sigma$ absorption band.}
\label{fig:AnalysisElectrons}
\end{figure}

For the sake of completeness we present in Fig.~\ref{fig:AnalysisElectrons} the results of analyzing 
a series of photoelectron images taken within the Rb $6s\Sigma$ band.
The slight shift to higher wave numbers of the spectral feature measured by detecting electrons 
with respect to ions [Fig.~\ref{fig:AnalysisElectrons} (a)] likely
results from contributions of ionized Rb$_2$ in the electron measurement.

The relative yields of photoelectrons out of the relaxed states 
$4d$, $5p_{3/2}$, and $5p_{1/2}$ are depicted in 
Fig.~\ref{fig:AnalysisElectrons} (b). Similarly to our previous 
measurements around the Rb $6p\Pi$ state, the relative populations
of the lowest detected levels increase as the laser is detuned below the droplet 
resonance. This change in relative populations likely reflects the variation of the monomer to dimer ratio. Changing relaxation rates into the various target electronic states of Rb due to droplet interactions may also contribute. 

The anisotropy parameters $\beta_2$ and $\beta_4$, which characterize 
the angular distribution of emitted electrons by two-photon 
ionization~\cite{Reid:2003}, are depicted in Fig.~\ref{fig:AnalysisElectrons} 
(c) and (d). The values of $\beta_2$ remain nearly constant within 
the accuracy of the measurement over the excitation spectrum. The 
$\beta_4$ values for the $5p$ states are roughly consistent with zero
for all laser wave numbers. This indicates vanishing  alignment of the 
electron orbitals as previously found for Rb 
$6p\Pi$ excitation~\cite{Fechner:2012}. However, the $4d$ orbital
appears to retain a certain degree of orbital alignment when exciting 
on the blue side of the Rb $6s\Sigma$ band. Likely, this is due to
faster desorption when exciting further up on the repulsive branch of 
the Rb-He$_N$ potential.

\section{\label{sec:Summary}Summary}
We have studied the desorption dynamics of the heavy alkali
metal atoms Rb and Cs off the surface of He nanodroplets, 
initiated by excitation to the perturbed $6s$ and $7s$ states,
respectively. As for Li and Na adatoms~\cite{Hernando:2012}, the
calculations reveal a complex response of the helium droplet to the impulsive perturbation induced by the excitation of the Rb and Cs adatoms. We find significant local 
deformations of the droplets and three distinct types of non-linear 
density waves which propagate through the droplets at different speeds. 
Nevertheless, both the measured and theoretically calculated mean 
kinetic energies of the desorbed atoms, which are in excellent 
agreement, can be modeled as a simple pseudodiatomic
direct photodissociation reaction driven by a highly repulsive
interaction. We find values of the effective mass of the He droplet 
interacting with Rb and Cs of about 10 and 13 He atoms,
respectively. Deviations from this simple model are found experimentally 
for the desorption dynamics of Rb on helium droplets excited to the $6p$
state.

The photoelectron spectra measured upon excitation to the perturbed $6s$ and $7s$ states evidence significant electronic relaxation of the desorbed Rb and Cs atoms into lower-lying states, at variance with analogous measurements using the light alkali species Li and Na attached to He droplets. While the ion and electron  measurements appear to be contradictory, possible correlations of the observed electrons with other ion signals can largely be ruled out. 

This puzzling issue will be further studied by measuring photoelectron spectra with femtosecond 
time-resolution in pump-probe experiments. Further theoretical work 
in this direction is also planned.

\begin{acknowledgement}

The authors gratefully acknowledge support by DGI, Spain (FEDER) under 
Grants No. FIS2011-28617-C02-01, by Generalitat de Catalunya (2009SGR1289),
and by the Deutsche Forschungsgemeinschaft. AL has been supported by 
the ME (Spain) FPI program, Grant No. BES-2012-057439.

\end{acknowledgement}



Animated views (mpeg-files) of the evolution of the helium density distributions upon excitation of rubidium and cesium adatoms are available as Supporting Information. This material is available free of charge via the Internet at http://pubs.acs.org.

\setkeys{acs}{usetitle = true}
\providecommand{\latin}[1]{#1}
\providecommand*\mcitethebibliography{\thebibliography}
\csname @ifundefined\endcsname{endmcitethebibliography}
  {\let\endmcitethebibliography\endthebibliography}{}

\setkeys{acs}{usetitle = true}

%
%

\begin{table}[p]
\begin{tabular}{c c c c c c c c c c c c c c c}
\hline \hline
Ak   &  \hspace{0.5 cm}    & m$_{Ak}$ (exp) & \hspace{0.5 cm} & $\eta$ (exp) & \hspace{0.5 cm} & $\eta$ (th) & \hspace{0.5 cm} & m$_{\mathrm{eff}}$ (exp) &\hspace{0.5 cm} &  m$_{\mathrm{eff}}$ (th) &\hspace{0.5 cm} &  $\langle r_e\rangle$ (\AA) &\hspace{0.5 cm} & $m_{\mathrm{eff}}$ (overlap) \\
\hline
Li    & & 6.94   & & 0.687  &  & 0.756 & & 15.2 & & 21.5 & & 5.35 & & 17.0  \\ 
Na & & 23.0 & & 0.516  &  & 0.583 & &  24.6 & & 32.2 & & 5.55 & & 22.1 \\ 
Rb        & & 85.5 & & 0.327  &  & 0.324 & & 41.9 & & 41.0  & & 6.54 & & 41.5\\
Cs        & & 132.9& & 0.281  &  & 0.273 & & 51.8 & &  50.5  & & 6.78 & & 53.8\\
[0.5ex] \hline 
\hline
\end{tabular}
\caption{\label{table1}
Characteristics of the experimental and theoretical kinetic energy distributions of the desorbed alkali atoms, see text for details. All masses are given in amu.
}
\end{table}

\end{document}